\documentclass[12pt]{article}
\usepackage{a4wide}
\usepackage{graphicx}
\usepackage{amsmath}
\usepackage{slashed}
\usepackage[retainorgcmds]{IEEEtrantools}
\usepackage{color}
\begin{document}
{\renewcommand{\thefootnote}{\fnsymbol{footnote}}
%\hfill  IGC--yy/m--n\\
\medskip
\begin{center}
{\LARGE Higher time derivatives in effective equations\\ of canonical quantum systems}\\
\vspace{1.5em}
Martin Bojowald\footnote{e-mail address: {\tt bojowald@gravity.psu.edu}},
Suddhasattwa Brahma\footnote{e-mail address: {\tt
    sxb1012@psu.edu}} and
Elliot Nelson\footnote{e-mail address: {\tt eln121@psu.edu}}
\\
\vspace{0.5em}
Institute for Gravitation and the Cosmos,\\
The Pennsylvania State
University,\\
104 Davey Lab, University Park, PA 16802, USA\\
\vspace{1.5em}
\end{center}
}

\setcounter{footnote}{0}

\begin{abstract}
  Quantum-corrected equations of motion generically contain higher time
  derivatives, computed here in the setting of canonically quantized
  systems. The main example in which detailed derivations are presented is a
  general anharmonic oscillator, but conclusions can be drawn also for systems
  in quantum gravity and cosmology.
\end{abstract}

\section{Introduction}

Quantum effects, generically, are non-local in time, captured in effective
actions or equations by correction terms containing time derivatives of higher
than second order. The path-integral formulation provides an intuitive
explanation: entire paths connecting two points, not just a local neighborhood
on a single trajectory, determine observable properties. In gravitational
theories, one expects quantum corrections with higher time derivatives for an
independent reason: The theory is interacting and therefore should receive
non-trivial quantum corrections. The only generally covariant extension of the
Einstein--Hilbert action by functionals of the metric is by higher powers and
contractions of space-time curvature tensors, most of which introduce time
derivatives of higher than second order in equations of motion.

Many properties of classical and quantum gravity are best described and
analyzed in a canonical formulation, especially when gauge issues play a
role. In canonical quantizations, however, it is not all too clear why and how
corrections with higher time derivatives should result. Equations of motion
with higher time derivatives imply additional degrees of freedom because more
initial data must be provided compared to usual second-order ones. In a
perturbative setting, the solution space does not increase in size because the
surplus solutions are not analytic in the perturbation parameter that
multiplies higher-derivative terms, and therefore must be discarded for
self-consistency \cite{Simon}. However, terms that contain higher time
derivatives do modify the classical solutions of second-order equations and
therefore the new degrees of freedom they come along with play an indirect
role. In canonical quantizations, however, one replaces the classical phase
space by a set of basic operators of the same number, without an obvious place
for new quantum degrees of freedom. A clear identification of such variables
together with a systematic procedure of deriving quantum corrections in which
they appear is of general interest, not just in quantum-gravity research where
higher-curvature corrections could be computed by such means.

Effective methods for canonical quantum systems do exist, using a systematic
analysis of quantum back-reaction of fluctuations and higher moments of a
state on the evolution of expectation values \cite{EffAc}. Applying this
scheme to anharmonic oscillators with potential
$V(q)=\frac{1}{2}m\omega^2q^2+U(q)$, it has been shown that results equivalent
to those of path-integral based low-energy effective actions \cite{EffAcQM}
are obtained. So far, these calculations, in \cite{EffAcQM} as well as
\cite{EffAc}, have been restricted to the first order in a semiclassical
expansion by $\hbar$ and second order in an adiabatic expansion, analogous to
a derivative expansion. To these orders, in
\begin{eqnarray} \label{GammaEff}
\Gamma_{{\rm eff}}[q(t)]&=&\int
{\rm d} t\Biggl(\frac{1}{2}\left(m+\frac{\hbar U'''(q)^2}
{32m^2\omega^5\left(1+U''(q)/m\omega^2\right)^{5/2}}\right)\dot
q^2\nonumber\\
&&
-\frac{1}{2}m\omega^2q^2-U(q)-\frac{\hbar\omega}{2}
\left(1+\frac{U''(q)}{m\omega^2}
\right)^{1/2}\Biggr)
\end{eqnarray}
one can see corrections by an effective quantum potential as well as a
correction to the particle mass, but no higher time derivatives result.

Nevertheless, the scheme provides a natural candidate for quantum
degrees of freedom analogous to new degrees of freedom in corrections
with higher-time derivatives \cite{Karpacz}: fluctuations and higher
moments of a state. As in perturbative higher-derivative theories,
these degrees of freedom play an indirect role when an adiabatic
expansion is used, because their equations of motion can be solved and
solutions can be inserted into equations for expectation values to
determine quantum corrections. In this article, we push the required
expansions to higher orders to compute several new correction terms
for the same systems, general anharmonic oscillators, and confirm that
higher time derivatives appear. These results are collected in
Sec.~\ref{s:Exp} and put together in Sec.~\ref{s:EOM}, after our
review of canonical effective techniques in Sec.~\ref{s:Anh}.

Looking at the details of our analysis, we will also be able to draw several
general conclusions about properties of effective canonical dynamics. These
statements, together with a general discussion of the relevance of our
findings for (loop) quantum cosmology, can be found in the concluding section
\ref{s:Conc}.

\section{Anharmonic oscillators}
\label{s:Anh}

The classical Hamiltonian for an anharmonic oscillator is given by:
\begin{equation} \label{H}
 H(q,p)= \frac{1}{2m} p^{2}+\frac{1}{2} m \omega^{2} q^{2}+U(q)
\end{equation}
with the particle mass $m$, harmonic frequency $\omega$, and an arbitrary
function $U(q)$ which, for many purposes such as quantum stability, is often
restricted to be bounded from below. We will mainly be thinking of a
polynomial $U(q)$ of higher than second order, whose total order is even if
boundedness from below is required. (Our effective equations will be
meaningful even when this condition is violated.) The frequency $\omega$ is
uniquely determined only if one requires $U(q)$ to have no quadratic
contribution.

The Hamiltonian (\ref{H}) can straightforwardly be quantized, without factor
ordering ambiguities. Quantum states $|\Psi\rangle(t)$ then  satisfy the
Schr\"odinger equation
\begin{equation}
 i\hbar\frac{\partial|\Psi\rangle}{\partial t}=\hat{H}|\Psi\rangle=
 \frac{1}{2m}\hat{p}^2|\Psi\rangle+ \frac{1}{2}m\omega^2\hat{q}^2|\Psi\rangle+
 U(\hat{q})|\Psi\rangle\,.
\end{equation}
This equation takes the form of a differential equation when a representation
of states, for instance as wave functions of $q$, is chosen. Effective
equations, however, are independent of this representation choice. (There may
be inequivalent representations not related unitarily, for instance on a
non-separable Hilbert space \cite{BohrQM}. In this case effective equations
would depend on which representation is chosen; see the example
\cite{BouncePert} in quantum cosmology.)

Instead of representations of wave functions, we use the general
evolution equation
\begin{equation} \label{dOdt}
 \frac{{\rm d}\langle\hat{O}\rangle}{{\rm d}t}=
 \frac{\langle[\hat{O},\hat{H}]\rangle}{i\hbar}
\end{equation}
for expectation values of observables $\hat{O}$. For a Hamiltonian as given
here, (\ref{dOdt}) applied to $\hat{q}$ and $\hat{p}$ gives rise to
Ehrenfest's equations
\begin{equation} \label{Ehrenfest}
 \frac{{\rm d}\langle\hat{q}\rangle}{{\rm d}t}=
   \frac{\langle\hat{p}\rangle}{m}\quad,\quad
 \frac{{\rm d}\langle\hat{p}\rangle}{{\rm d}t}= -\langle V'(\hat{q})\rangle\,,
\end{equation}
resembling the classical ones but also exhibiting quantum corrections
in the force term $-\langle V'(\hat{q})\rangle$ compared to
$-V'(\langle\hat{q}\rangle)$. Canonical effective equations compute
the difference of $-\langle V'(\hat{q})\rangle$ and
$-V'(\langle\hat{q}\rangle)$ in a systematic way. (In \cite{Hepp},
these equations are used to prove that quantum mechanics has the
correct classical limit for $\hbar\to0$.)

Unless the potential is at most quadratic, Eqs.~(\ref{Ehrenfest}) do not
provide a closed set of equations that could be solved for the expectation
values, starting from some initial values. A cubic term $\lambda q^3$ in the
potential, for instance, gives rise to $-3\lambda \langle \hat{q}^2\rangle=
-3\lambda (\langle\hat{q}\rangle^2+ (\Delta q)^2)$. The second term is a
quantum correction to the classical force $-3\lambda q^2$, but it depends on
the position fluctuation $\Delta q$ which is independent of the expectation
value $\langle \hat{q}\rangle$. Ehrenfest's equations therefore cannot be
solved in this case, unless one already knows how the quantum state or at
least its position fluctuation evolves.

To provide a complete set of equations, effective techniques enlarge the set
of Ehrenfest's equations by deriving differential equations for fluctuations
and all moments
\begin{equation}\label{Moments}
 \tilde{G}^{a,n}:=\left\langle (\hat{q}-\left\langle \hat{q}\right\rangle
   )^{n-a}(\hat{p}-\left\langle \hat{p}\right\rangle
   )^{a}\right\rangle_{\rm Weyl}
\end{equation}
where the subscript ``Weyl'' indicates Weyl (or totally symmetric)
ordering. Since these variables are defined by expectation values, equations
of motion for them can be derived from the same equation (\ref{dOdt}) as used
earlier. Moreover, these variables provide infinitely many quantum degrees of
freedom independent of expectation values, just the degrees of freedom which,
at least qualitatively, should be related to implications of higher time
derivatives.

Unlike expectation values, moments cannot take arbitrary values. For $n=2$,
they must satisfy the familiar uncertainty relation
\begin{equation} \label{Uncert}
 \tilde{G}^{0,2}\tilde{G}^{2,2}- (\tilde{G}^{1,2})^2\geq \frac{\hbar^2}{4}
\end{equation}
and there are analogous, but less familiar relations at higher orders. All
these inequalities follow from the Schwarz inequality. Only if
they are obeyed can the moments correspond to a state. If they are, the state
may be pure or mixed; if a selection of pure states is required, additional
conditions must be imposed. (See also the examples in \cite{Springer}.)

Instead of calculating all the commutators required for equations of motion
(\ref{dOdt}) of moments, it is usually more straightforward to take a
phase-space point of view. We first consider the right-hand side of
(\ref{dOdt}), which in classical mechanics would be the Poisson bracket of $O$
with the Hamiltonian $H$. This comparison motivates the definition of a
Poisson bracket between expectation values of arbitrary operators,
\begin{equation} \label{Poisson}
 \{\langle\hat{A}\rangle,\langle\hat{B}\rangle\}:=
 \frac{\langle[\hat{A},\hat{B}]\rangle}{i\hbar}\,.
\end{equation}
This bracket is antisymmetric, linear and satisfies the Jacobi identity by
virtue of those properties realized for the commutator. If we extend it to
products of expectation values by using the Leibniz rule, we obtain a
well-defined Poisson bracket on the quantum phase space, whose elements are
states parameterized by their expectation values of $\hat{q}$ and $\hat{p}$
together with all moments. (The quantum phase space is symplectic. However, in
effective equations one truncates the system at finite orders of $\hbar$,
which translates to finite orders of the moments. The Poisson tensor remains
well-defined on these subspaces, but in general is no longer
invertible. Effective equations therefore cannot be described by symplectic
techniques.)

Combining (\ref{dOdt}) and (\ref{Poisson}), the Schr\"odinger flow is
described equivalently by a Hamiltonian flow on the quantum phase space,
generated by the quantum Hamiltonian
$H_Q(\langle\hat{q}\rangle,\langle\hat{p}\rangle, \tilde{G}^{a,n}):=
\langle\hat{H}\rangle_{\langle\hat{q}\rangle,\langle\hat{p}\rangle,
  \tilde{G}^{a,n}}$. The subscript indicates that the expectation value
is taken in a state characterized by the values
$\langle\hat{q}\rangle,\langle\hat{p}\rangle, \tilde{G}^{a,n}$; the result
then defines the value of the quantum Hamiltonian at the quantum phase-space
point corresponding to the same state.

As a function of expectation values and moments, quantum Hamiltonians can be
computed by writing $\langle H(\hat{q},\hat{p})\rangle= \langle
H(\langle\hat{q}\rangle+ (\hat{q}-\langle\hat{q}\rangle),
\langle\hat{p}\rangle+ (\hat{p}-\langle\hat{p}\rangle))\rangle$ and expanding
in the ``small'' quantities $\hat{q}-\langle\hat{q}\rangle$ and
$\hat{p}-\langle\hat{p}\rangle$. This formal expansion is a shortcut for a
direct computation of the expectation value. Inserting a Taylor expansion and
assuming $H(\hat{q},\hat{p})$ to be Weyl ordered (in case there is any
factor-ordering choice), we obtain
\begin{equation}
H_{Q}:=\langle H(\hat{q},\hat{p})\rangle=
\sum_{n=0}^{\infty}\sum_{a=0}^{n}\frac{1}{n!}\binom{n}{a} \frac{\partial^n
  H(q,p)}{\partial p^a\partial q^{n-a}}\tilde{G}^{a,n}
\end{equation}
with the moments (\ref{Moments}). (Here and from now on we identify
$q:=\langle\hat{q}\rangle$ and $p:=\langle\hat{p}\rangle$ to simplify our
notation.)

The moments may be written in dimensionless form as
$G^{a,n}=\hbar^{-n/2}(m\omega)^{n/2-a}\tilde{G}^{a,n}$ to facilitate future
expansions. In terms of these dimensionless variables, we obtain, for the given
anharmonic oscillator, the quantum Hamiltonian as:
\begin{equation}
H_{Q}=\frac{1}{2m} p^{2}+\frac{1}{2} m \omega^{2}
q^{2}+U(q)+\frac{\hbar\omega}{2}(G^{0,2}+G^{2,2})+
\sum_{n=2}^{\infty}\frac{1}{n!}(\hbar/m\omega)^{n/2}U^{(n)}(q)G^{0,n}\,.
\end{equation}
The first quantum correction, depending only on fluctuations but not on
expectation values, is a zero-point energy. The sum, on the other hand,
contains products of expectation values and moments if there is an anharmonic
potential, and therefore describes the coupling between quantum variables and
expectation values, or quantum back-reaction. (Note that the moments vanish
identically if $n=1$. The sum therefore starts at $n=2$.)

If we look at the equations of motion for expectation values and moments
generated by the quantum Hamiltonian \cite{EffAc},
\begin{eqnarray}
\dot{q}&=&\{q,H_{Q}\}=\frac{p}{m}\nonumber\\
\dot{p}&=&\{p,H_{Q}\}=-m\omega^{2}q-U'(q)-
\sum_{n=2}^{\infty}\frac{1}{n!}(\hbar/m\omega)^{n/2}U^{(n+1)}(q)G^{0,n}
\label{pdot}\\
\dot{G}^{a,n}&=&-a\omega G^{a-1,n}+(n-a)\omega G^{a+1,n} -
\frac{U''(q)a}{m\omega} G^{a-1,n} \label{Gdot}\\
\nonumber&&+
\frac{\sqrt{\hbar}aU'''(q)}{2(m\omega)^{3/2}}
G^{a-1,n-1} G^{0,2}
+\frac{\hbar
  aU^{''''}(q)}{3!(m\omega)^2} G^{a-1,n-1}G^{0,3} \\
\nonumber&&
-\frac{a}{2}\left(
\frac{\sqrt{\hbar}U'''(q)}{(m\omega)^{3/2}}
G^{a-1,n+1}+\frac{\hbar
U^{''''}(q)}{3(m\omega)^2}G^{a-1,n+2}\right)+\cdots
\end{eqnarray}
(not all terms are written in the last equation) we can already see
that moments are related to higher time derivatives: Eq.~(\ref{pdot})
can be interpreted as identifying an infinite linear combination of
moments with the second derivative of $q$, in a way that also depends
on $q$ itself. Taking further time derivatives of the whole equation
(\ref{pdot}) and inserting (\ref{Gdot}) relates different combinations
of the $G^{a,n}$ (no longer linear) to time derivatives of $q$ of
higher than second order. It is therefore clear that moments in the
canonical setting play the role of higher time derivatives in a
Lagrangian one. But so far the identification is not very direct, and
the equations we obtain for higher time derivatives in terms of
moments are difficult to invert. In the rest of this paper, we work
out a systematic method, using two expansions as in \cite{EffAc}, to
write (\ref{pdot}) as an equation corrected by higher-derivative
terms, eliminating the moments.

\section{Semiclassical and Adiabatic Expansions}
\label{s:Exp}

Our first expansion is a semiclassical one. In a semiclassical state, the
moments by definition obey the $\hbar$ierarchy
$\tilde{G}^{a,n}=O(\hbar^{n/2})$ so that an expansion by $\hbar$ to a given
finite order makes use of only finitely many moments. Thanks to the definition
of dimensionless variables
$G^{a,n}=\hbar^{-n/2}(m\omega)^{n/2-a}\tilde{G}^{a,n}$, suitable powers of
$\hbar$ already appear as factors in equations of motion such as (\ref{pdot}),
and we only need to truncate the sum.

Although the leading order of $\hbar$ is split off the moments when
using dimensionless variables, each moment, as a solution of
(\ref{Gdot}), could still have higher-order corrections in
$\hbar$. For full generality, we therefore make these terms explicit
by expanding by powers of $\sqrt{\hbar}$: $G^{a,n}=\sum_e G^{a,n}_e
\hbar^{e/2}$. The coefficients $G^{a,n}_e$ are then independent of
$\hbar$. Some features of this expansion may look unfamiliar, which we
explain by two comments:
\begin{itemize}
\item As written explicitly, we should expect half-integer orders in $\hbar$
  because a moment of order $n$ behaves as $O(\hbar^{n/2})$. If all odd-order
  moments vanish, which is the case in a large subclass of semiclassical
  states, only integer orders appear, as naively expected. Such an assumption
  can always be made for an initial state, but non-trivial quantum
  back-reaction can easily generate non-vanishing odd-order moments. (See
  for instance the example in \cite{HigherMoments}.)
\item Since $\hbar$ has non-trivial dimensions, the coefficients
  $G_e^{a,n}$ have different dimensions for different $e$. If this
  feature is unwanted, one can use coefficients or parameters in the
  anharmonicity potential $U(q)$, together with $m$ and $\omega$, to
  define a parameter $L$ of the same dimensions as $\hbar$ and expand
  by $\sqrt{\hbar/L}$. The parameters of the harmonic-oscillator
  Hamiltonian have already been used to absorb the dimensions of
  $\tilde{G}^{a,n}$, and they do not allow a combination with the
  dimensions of $\hbar$. In the harmonic case, an expansion by
  $\sqrt{\hbar/L}$ cannot be done, and it is not necessary because the
  equations of motion for moments can then be solved exactly, showing
  that each moment $\tilde{G}^{a,n}$ is exactly proportional to
  $\hbar^{n/2}$ and stays so at all times. This property is no longer
  realized for anharmonic oscillators, but then there are additional
  parameters in the potential that can be used to define a suitable
  $L$. We refrain from doing so here because we work with a general
  anharmonicity $U(q)$. Its derivatives then provide the correct
  dimensions.
\end{itemize}

With the semiclassical expansion in $\hbar$, the equations of motion for the
moments partially decouple.
At $O(\hbar^0)$ and $O(\hbar^{1/2})$, we have
\begin{eqnarray}
\dot{G}^{a,n}_0 & = & -a\omega G^{a-1,n}_0+(n-a)\omega G^{a+1,n}_0 -
\frac{U''(q)a}{m\omega} G^{a-1,n}_0
\label{Gdot0}
\\
\dot{G}^{a,n}_1 & = & -a\omega G^{a-1,n}_1+(n-a)\omega G^{a+1,n}_1 -
\frac{U''(q)a}{m\omega} G^{a-1,n}_1 +
\frac{U'''(q)a}{2(m\omega)^{3/2}}G^{0,2}_0 G^{a-1,n-1}_0 \nonumber\\ && \: -
\frac{U'''(q)a}{2(m\omega)^{3/2}} \left(G^{a-1,n+1}_0 - \frac{(a-1)(a-2)}{12}
G^{a-3,n-3}_0\right)
\label{Gdot1}
\end{eqnarray}
which we could try to solve order by order. The equations for
$G_0^{a,n}$ couple only the $n+1$ moments at fixed order $n$ and are
linear in the moments. The equations for $G_1^{a,n}$ (and similarly
for higher orders in $\hbar$) are inhomogeneous but also contain only
moments $G_1^{a,n}$ of the same order $n$, as well as non-linear
inhomogeneous terms of different orders.

By the semiclassical expansion, the infinitely coupled original system has
been reduced to finitely coupled subsets.  In principle one could solve this
system order by order, but since coefficients of the differential equations
also depend on $q$ for a non-trivial anharmonicity, they are difficult to
solve explicitly. We therefore make use of a second expansion, an adiabatic
one, which reduces the differential equations to algebraic ones. This
approximation will also be crucial to bring out the nature of moments as
higher time derivatives. (In the conclusions we will comment on the nature of
moments in regimes in which no adiabatic expansion is possible.)

The adiabatic expansion is defined by replacing all time derivatives in
equations of motion for moments ($q$ and $p$ are not assumed to change just
adiabatically) by ${\rm d}/{\rm d}t\rightarrow \lambda{\rm d}/{\rm d}t$,
expanding all coefficients $G_e^{a,n}=\sum_{i=1}^{\infty}
G_{e,i}^{a,n}\lambda^i$ in $\lambda$, solving equations order by order in
$\lambda$, and setting $\lambda=1$ in the end. The expansion is formal because
there is no guarantee that the series converges for $\lambda=1$. Moreover, the
parameter $\lambda$, unlike $\hbar$ in the semiclassical expansion, has no
physical meaning. As we will see later, the procedure rather serves as a
systematic way of organizing the appearance of derivatives of different
orders.

At zeroth order of the adiabatic approximation of $\dot{G}^{a,n}=
\{G^{a,n},H_Q\}$, we have equations
\begin{equation}
0=\{G^{a,n}_{e,0},H_Q\}
\end{equation}
which are algebraic rather than differential. At higher orders,
\begin{equation} \label{Gdoti}
\dot{G}^{a,n}_{e,i}=\{G^{a,n}_{e,i+1},H_Q\}
\end{equation}
contains time derivatives, but if we proceed order by order, we can assume
that $G_{e,i}^{a,n}$ has already been solved for, starting with the algebraic
equation for $G_{e,0}^{a,n}$. With the time dependence on the left-hand side
of (\ref{Gdoti}) known, the equation again reduces to an algebraic one for
$G^{a,n}_{e,i+1}$. Proceeding order by order, the main equations to be solved
are algebraic. (Some differential consistency conditions also arise, as we
will see explicitly.)

Combining both expansions, our moments read
\begin{equation}
G^{a,n}=\sum_e \sum_i G^{a,n}_{e,i} \hbar^{e/2} \lambda^i\,.
\end{equation}
Equations to be solved for the coefficients $G_{e,i}^{a,n}$ show both
advantages noted above: They split into finitely coupled sets, and are
algebraic.

We now proceed to computing explicit solutions to several orders.  But first
we interject a comment on our designation of $\hbar$-orders to avoid
confusion. For most of this paper, we will be looking at the moment equations
(\ref{Gdot}) and their solutions, and therefore speak of the
$O(\hbar^0)$-order when all $\hbar$-terms are dropped, of the
$O(\hbar^{1/2})$-order when terms linear in $\sqrt{\hbar}$ are kept, and so
on. These orders give us the relevant information about state properties at
the corresponding orders. However, if we use these moment solutions to compute
correction terms for effective equations of expectation values, inserting the
moments in (\ref{pdot}) as we will do in the end, the $\hbar$-orders shift
because (\ref{pdot}) contains explicit factors of $\hbar$. Somewhat
counter-intuitively, even the $O(\hbar^0)$-order in the moments will then
contribute to non-trivial quantum corrections. This intermingling of the
orders cannot be avoided because equations of motion for different variables
--- expectation values or moments of different orders $n$ --- have their own
arrangements of $\hbar$-terms.

\subsection{Adiabatic Approximation at $O(\hbar^0)$ in the Moments}

At zeroth order in the adiabatic approximation, we can ignore all time
dependence: $\{G^{a,n}_{e,0},H_Q\}=0$.  At zeroth order also in the
$\sqrt{\hbar}$ expansion, we have from (\ref{Gdot0})
\begin{equation}
0=-a\omega G^{a-1,n}_{0,0}+(n-a)\omega G^{a+1,n}_{0,0} -
\frac{U''(q)a}{m\omega} G^{a-1,n}_{0,0}
\label{18}
\end{equation}
for $0\leq a\leq n$, which gives a solution of the form
\begin{equation} \label{00solutionC}
G^{a,n}_{0,0}=C_n \frac{(n-a)!a!}{((n-a)/2)!(a/2)!}
\left(1+\frac{U''(q)}{m\omega^2}\right)^{(2a-n)/4}
\end{equation}
with some coefficients $C_n$ if both $n$ and $a$ are even. Otherwise,
$G_{0,0}^{a,n}=0$. (For odd $a$, Eq.~(\ref{18}) used with $a=0$ implies that
$G^{1,n}_{0,0}=0$, which upon recurrence to $a=2k+1$ with integer $k$ implies
that $G^{a,n}_{0,0}=0$ for odd $a$, no matter whether $n$ is even or odd. For
$n$ odd and $a$ even in $G_{0,0}^{a,n}$, Eq.~(\ref{18}) evaluated for $a=n$ is
meaningful, with a zero value implied for $G^{n+1,n}_{0,0}$, only if all
$G^{a,n}_{0,0}$ with odd $n$ and even $a$ vanish.)

Only the values (\ref{00solutionC}) with even $n$ and $a$ can be
non-zero. Based on the zeroth-order equation (\ref{Gdot0}) alone, the
$C_n$ could depend on $q$ as well, but this possibility is ruled out
by a consistency condition obtained at first adiabatic order
\cite{EffAc,Karpacz}. The values of $C_n$ in general effective
equations remain free (provided the resulting moments satisfy the
uncertainty relation) and parameterize different choices of adiabatic
states. A prominent choice is the anharmonic vacuum, whose values can
be obtained by requiring that the harmonic limit $U(q)=0$ provides the
known moments of the harmonic-oscillator ground state. The
$C_n=2^{-n}$ \cite{EffAc} are then fixed, and we have
\begin{equation}
G^{a,n}_{0,0}=\frac{(n-a)!a!}{2^n ((n-a)/2)!(a/2)!}
\left(1+\frac{U''(q)}{m\omega^2}\right)^{(2a-n)/4}
\label{00solution}
\end{equation}
for even $a$ and $n$.

As recalled here, the derivation of (\ref{00solution}) with its
precise coefficient requires additional assumptions about the initial
values of the moments, or about the kind of states whose evolution is
considered. General effective equations are not unique owing to the
dependence on classes of states described by them. The usual
low-energy effective action (\ref{GammaEff}) is parameter-free only
because it refers to a specific regime of states near the interacting
vacuum, as indicated by the qualifier ``low-energy.'' The underlying
conditions are tantamount to requiring the moments to agree with those
of the harmonic-oscillator ground state when $U(q)=0$, and indeed
canonical effective equations with this choice are equivalent to
equations of motion that follow from the low-energy effective action
\cite{EffAc}.  The solutions for moments then amount to expanding
around the adiabatic vacuum state of the anharmonic system. Since our
results build on (\ref{00solution}), we will be dealing with the same
states, but to higher orders in the semiclassical and adiabatic
expansions.

\subsubsection{Solutions at zeroth and first adiabatic order}

Using zeroth-order solutions in $\dot{G}^{a,n}_{0,0}=\{G^{a,n}_{0,1},H_Q\}$,
the equation of motion at first order in $\lambda$ and zeroth order in
$\sqrt{\hbar}$, we have, for odd $a$ or $n$,
\begin{equation}
\dot{G}^{a,n}_{0,0}=0=-a\omega G^{a-1,n}_{0,1}+(n-a)\omega G^{a+1,n}_{0,1} -
\frac{U''(q)a}{m\omega} G^{a-1,n}_{0,1}
\label{19}
\end{equation}
But this equation is identical to \eqref{18}, so we have the same solution,
namely zero, for odd $n$. This pattern continues to all orders in the
adiabatic approximation:
\begin{equation}
G^{a,n}_{0,i}=0 \ \ \text{for odd} \ n.
\label{oddn}
\end{equation}

For even $n$, however, solutions change with progressing adiabatic
order.  We can still use \eqref{19} for $a$ odd and $n$ even,
describing solutions $G^{a,n}_{0,1}$ with even $a$ and $n$.  Again,
the equation is identical to \eqref{18}, solved by \eqref{00solutionC}
with new coefficients $C_n'$ instead of $C_n$. If we match with the
harmonic-oscillator ground state, we must require its values for the
moments $G_{0,0}^{a,n}+G_{0,1}^{a,n}$ to the present order, so that
$C_n+C_n'=2^{-n}$. Only this combination of $C_n$ and $C_n'$ appears
in equations of motion to first adiabatic order, and therefore it is
not necessary (nor possible) to determine both coefficients
independently. Since we already fixed $C_n$, we keep this value as
well as $G_{0,0}^{a,n}$ as in (\ref{00solution}). This choice implies
$C_n'=0$ and therefore
\begin{equation}
G^{a,n}_{0,1}=0 \ \ \text{for even} \ a \ \text{and} \ n\,.
\label{01eveneven}
\end{equation}

For moments of odd $a$ and even $n$, finally, the solution of
(\ref{19}) (with odd $a$ inserted) is different.  In this case, the
time derivative of \eqref{00solution} becomes the left-hand side of
the first-order adiabatic equation of motion:
\begin{eqnarray}
&&\frac{(n-a)!a!}{2^n
  ((n-a)/2)!(a/2)!}\frac{2a-n}{4}\frac{U'''(q)\dot{q}}{m\omega^2}
\left(1+\frac{U''(q)}{m\omega^2}\right)^{(2a-n-4)/4}\nonumber\\ \quad
&=&-a\omega G^{a-1,n}_{0,1}+(n-a)\omega G^{a+1,n}_{0,1} -
\frac{U''(q)a}{m\omega} G^{a-1,n}_{0,1}
\label{01EOM}
\end{eqnarray}
We can solve this immediately for $a=n$, and then substitute the
result in to solve for the case $a=n-2$, and so on.  The general solution is
\begin{equation}
G^{a,n}_{0,1}=C_{a,n}\frac{U'''(q)\dot{q}}{m\omega^3}
\left(1+\frac{U''(q)}{m\omega^2}\right)^{(2a-n-6)/4}
\ \text{for odd} \ a \ \text{and even} \ n
\label{01oddeven}
\end{equation}
where the $C_{a,n}$ are dimensionless prefactors. In particular, while the
zeroth adiabatic order does not allow quantum correlations, they may appear at
first adiabatic order, for instance by $G^{1,2}\not=0$.

In the harmonic limit, (\ref{01oddeven}) vanishes identically, and
therefore the $C_{a,n}$ are not restricted by the requirement of
perturbing around the harmonic ground state. Instead, the $C_{a,n}$
are fully determined by the adiabatic equations.  For $a=n$ in
(\ref{01EOM}), we have $C_{n-1,n}=-2^{-(n+2)}n!/(n/2)!$. Plugging
\eqref{01oddeven} into \eqref{01EOM} we find
\begin{equation}\label{can}
C_{a-1,n}=\frac{n-a}{a}C_{a+1,n} -
\frac{(n-a)!(a-1)!}{2^{n+2}((n-a)/2)!(a/2)!}(2a-n) \,,
\end{equation}
a recurrence relation solved by the general expression
\begin{eqnarray}
C_{a-1,n} &=& -
\frac{(n-a)!(a-1)!}{2^{n+2}((n-a)/2)!(a/2)!}(2a-n)\\
&&
-2^{-n-2}\sum_{b=0}^{(n-a-2)/2}
\left[\prod_{c=0}^{b}\frac{n-(a+2c)}{a+2c}\right]
\frac{(n-a_b')!(a_b'-1)!}{((n-a_b')/2)!(a_b'/2)!}(2a_b'-n) \nonumber
\end{eqnarray}
for even $a$, where $a_b'=a+2(b+1)$.

To summarize, at first adiabatic order only moments with odd $a$ and
even $n$ change, depending on the time derivative of $q$ in
(\ref{01oddeven}).

\subsubsection{Second Adiabatic Order}

At second order in $\lambda$ and zeroth order in $\hbar$, the equation
of motion is
\begin{equation} \label{G0102}
\dot{G}^{a,n}_{0,1} =
-a\omega\left(1+\frac{U''(q)}{m\omega^2}\right)G^{a-1,n}_{0,2}+(n-a)\omega
G^{a+1,n}_{0,2}\,.
\end{equation}
For odd $n$ the solution is zero by \eqref{oddn}, and even $a$ and $n$
in the equation of motion leads to $G^{a,n}_{0,2}$ of the form
(\ref{00solutionC}) for odd $a$ and even $n$, again with new
coefficients $C_n$. For these moments to vanish in the harmonic limit,
we have $G_{0,2}^{a,n}=0$ for odd $a$.

For odd $a$ and even $n$, the left-hand side is given by the time
derivative of equation \eqref{01oddeven}.  Substituting in this result
and rearranging slightly, we have
\begin{eqnarray}
G^{a+1,n}_{0,2}=&&\frac{a}{n-a}
\left(1+\frac{U''(q)}{m\omega^2}\right)G^{a-1,n}_{0,2}\nonumber\\
&&+\frac{C_{a,n}}{(n-a)m\omega^4}\bigg((U'''(q)\ddot{q}
+U''''(q)\dot{q}^2)\left(1+\frac{U''(q)}{m\omega^2}\right)^{(2a-n-6)/4}
\nonumber \\
&&+(U'''(q)\dot{q})^2\frac{2a-n-6}{4m\omega^2}
\left(1+\frac{U''(q)}{m\omega^2}\right)^{(2a-n-10)/4}\bigg) \,.
\label{02eveneven}
\end{eqnarray}
The form of this equation suggests an ansatz
\begin{eqnarray}
G^{a,n}_{0,2}&&=A_{a,n}\left(\frac{1}{m\omega^4}(U'''(q)\ddot{q}
+U''''(q)\dot{q}^2)\left(1+\frac{U''(q)}{m\omega^2}\right)^{(2a-n-8)/4}\right)
\nonumber \\
&&+B_{a,n}\left(\frac{1}{4m^2\omega^6}(U'''(q)\dot{q})^2
\left(1+\frac{U''(q)}{m\omega^2}\right)^{(2a-n-12)/4}\right) \nonumber\\
&&+{{n/2}\choose{a/2}}{{n}\choose{a}}^{-1}
\left(1+\frac{U''(q)}{m\omega^2}\right)^{a/2}G^{0,n}_{0,2}
\label{32}
\end{eqnarray}
where $A_{a,n}$ and $B_{a,n}$ are dimensionless coefficients
determined by $a$ and $n$, and the final term is motivated by the
solution at zeroth order in $\lambda$, generated by the first term in
\eqref{02eveneven}.  Using this ansatz in \eqref{02eveneven}, we find
recursion relations for $A_{a,n}$ and $B_{a,n}$:
\begin{eqnarray}
A_{a+1,n}&=&\frac{C_{a,n}}{n-a}+\frac{a}{n-a}A_{a-1,n} \\
B_{a+1,n}&=&\frac{C_{a,n}(2a-n-6)}{n-a}+\frac{a}{n-a}B_{a-1,n}
\end{eqnarray}
for odd $a$.  Consistency of (\ref{32}) requires that
$A_{0,n}=B_{0,n}=0$, which gives $A_{2,n}=\frac{C_{1,n}}{n-1}$ and
$B_{2,n}=-\frac{n+4}{n-1}C_{1,n}$.  The recursion relations, with
these initial values, give
\begin{equation}
A_{a+1,n}=\frac{C_{a,n}}{n-a}+\sum_{b=0}^{(a-3)/2}
\left(\prod_{c=0}^{b}\frac{a-2c}{n-(a-2c)}\right)\frac{C_{a-2(b+1),n}}{n-(a-2(b+1))}
\end{equation}
for odd $a\geq3$, where the $C_{a,n}$ are given in \eqref{can}.  The
expression for the $B_{a,n}$ is similar, the only change being that
$C_{a,n}$ is replaced by $(2a-n-6)C_{a,n}$ in each term.

To fully determine the moments, which are thus far given in terms of
$G^{0,n}_{0,2}$, we need a condition from third order.  Since we are still at
$O(\hbar^0)$, the third-order equation of motion has the same form, so the
condition is given by \cite{Karpacz}
\begin{equation}
\sum_{\text{even } a}
{{n/2}\choose{a/2}}\left(1+\frac{U''(q)}{m\omega^2}\right)^{(n-a)/2}
\dot{G}^{a,n}_{0,2} =0
\label{constraint}
\end{equation}
for even $n$.  From \eqref{32} we see that this is a complicated differential
equation for $G^{0,n}_{0,2}$.  Given \eqref{32}, this condition suggests an
ansatz for $G^{0,n}_{0,2}$ of the form
\begin{equation}
G^{0,n}_{0,2}=A'_n\frac{1}{m\omega^4}(U'''(q)\ddot{q}
+U''''(q)\dot{q}^2)\left(1+\frac{U''(q)}{m\omega^2}\right)^r
+B'_n\frac{1}{4m^2\omega^6}(U'''(q)\dot{q})^2
\left(1+\frac{U''(q)}{m\omega^2}\right)^s
\label{constraintansatz}
\end{equation}
where $r$, $s$, $A'_n$, and $B'_n$ are some undetermined constants.
(Given a differential equation (\ref{constraint}), we expect one free
parameter, which would be multiplying all $G_{0,2}^{a,n}$. However,
since we are also solving the recurrence relation (\ref{32}), whose
first two terms are fixed, imposing the consistency condition
(\ref{constraint}) will not leave any free parameters.)

Substituting \eqref{32} into \eqref{constraint} with \eqref{constraintansatz},
differentiating, and moving things around a bit, we have that
\begin{eqnarray}
&& \frac{1}{m\omega^4} \sum_a \left\{ {{n/2}\choose{a/2}}A_{a,n} \left[
  \frac{\text{d}\left(U'''\ddot{q}
+U''''\dot{q}^2\right)}{\text{d}t}X^{\frac{n-8}{4}}
  +
  \frac{2a-n-8}{4}\frac{U'''\dot{q}}{m\omega^2}
\left(U'''\ddot{q}+U''''\dot{q}^2\right)X^{\frac{n-12}{4}}
\right]\right. \nonumber \\
&&+\left. A'_n {{n/2}\choose{a/2}}^2 {{n}\choose{a}}^{-1}
\left[
  \frac{\text{d}\left(U'''\ddot{q}+U''''\dot{q}^2\right)}{\text{d}t}
X^{r+\frac{n}{2}}
  +
  \left(r+\frac{a}{2}\right)\frac{U'''\dot{q}}{m\omega^2}
\left(U'''\ddot{q}+U''''\dot{q}^2\right)X^{r+\frac{n}{2}-1}
\right]\right\} \nonumber \\
&&+\frac{1}{4m^2\omega^6} \sum_a\left\{ {{n/2}\choose{a/2}}B_{a,n} \left[
  2U'''\dot{q}\left(U'''\ddot{q}+U''''\dot{q}^2\right)X^{\frac{n-12}{4}} +
  \frac{2a-n-12}{4}\frac{U'''\dot{q}}{m\omega^2}\left(U'''\dot{q}\right)^2
  X^{\frac{n-16}{4}}
\right]\right. \nonumber \\
&&+\left. B'_n  {{n/2}\choose{a/2}}^2
{{n}\choose{a}}^{-1} \left[
  2U'''\dot{q}\left(U'''\ddot{q}+U''''\dot{q}^2\right)X^{s+\frac{n}{2}} +
  \left(s+\frac{a}{2}\right)\frac{U'''\dot{q}}{m\omega^2}
\left(U'''\dot{q}\right)^2
  X^{s+\frac{n}{2}-1} \right]\right\}
\label{long}
\end{eqnarray}
must vanish, where $X=1+U''/m\omega^2$.  By inspection, we see that for
$r=-(n+8)/4$ and $s=-(n+12)/4$, terms involving the same expressions with $q$
and its derivatives also have the same power of $X$.  This leaves only the
numerical coefficients to be fixed.  Only the first terms in the first two
lines are proportional to
$X^{(n-8)/4}\text{d}(U'''\ddot{q}+U''''\dot{q}^2)/\text{d}t$; generically,
these terms must add to zero separately, which allows us to solve
for
\begin{equation}
A'_n=-\frac{\sum_a {{n/2}\choose{a/2}} A_{a,n}}{\sum_a{{n/2}\choose{a/2}}^2
  {{n}\choose{a}}^{-1}} \,.
\end{equation}
Similarly, only the last terms in the last two lines are proportional to
$(U'''\dot{q})^3 X^{(n-16)/4}$, the remaining terms being proportional
to $U'''\dot{q}\left(U'''\ddot{q}+U''''\dot{q}^2\right)X^{(n-12)/4}$, so
we can solve for $B'_n$ in similar fashion:
\begin{equation}
B'_n=-\frac{\sum_a {{n/2}\choose{a/2}} B_{a,n}
  (2a-n-12)}{\sum_a{{n/2}\choose{a/2}}^2 {{n}\choose{a}}^{-1} (2a-n-12)} \,.
\end{equation}

To confirm that these expressions are indeed valid, we need to check that the
remaining terms in \eqref{long} vanish.  Substituting in our expressions for
$A'_n$ and $B'_n$, factoring out common quantities, and making use of the fact
that $\sum_a {{n/2}\choose{a/2}}^2 {{n}\choose{a}}^{-1}(2a-n)=0$, we find that
\begin{equation}
\text{remaining terms }\propto \ \sum_a
{{n/2}\choose{a/2}}(2a-n)\left(6A_{a,n}+B_{a,n}\right)\,.
\end{equation}
We have checked that this expression vanishes for $n=2,4,6$, confirming the
solution
\begin{eqnarray}\label{G0}
G^{0,n}_{0,2}&=&\frac{A'_n}{m\omega^4}(U'''(q)\ddot{q}+U''''(q)\dot{q}^2)
\left(1+\frac{U''(q)}{m\omega^2}\right)^{-(n+8)/4} \\
&&+\frac{B'_n}{4m^2\omega^6}(U'''(q)\dot{q})^2
\left(1+\frac{U''(q)}{m\omega^2}\right)^{-(n+12)/4}\nonumber
\end{eqnarray}
at least to these orders, where $A'_n$ and $B'_n$ are given above.  This
expression also reduces to the solution for $G^{0,2}_{0,2}$ given in
\cite{Karpacz}, with the correct coefficients $A'_2=1/16$ and $B'_2=-5/16$,
which can be checked using the earlier expressions for $A_{a,n}$, $B_{a,n}$,
and $C_{a,n}$. The solution for $G_{0,2}^{0,n}$ does not modify the
ground-state condition for $C_n$ in (\ref{00solutionC}) because it
automatically vanishes in the harmonic limit $U(q)=0$. Instead, the
coefficients $A_n'$ and $B_n'$ are completely fixed without a choice of state.

\subsection{Adiabatic Approximation at $O(\sqrt{\hbar})$ in the Moments}

Starting with the first order in $\sqrt{\hbar}$, we need to consider different
adiabatic orders in separation.

\subsubsection{Zeroth Adiabatic Order}
\label{sec:10}

At leading order in the adiabatic approximation, $\{G^{a,n}_{e,0},H_Q\}=0$.
At $O(\sqrt{\hbar})$, that is, $e=1$, this gives (from Eq. \eqref{Gdot1})
\begin{eqnarray}
0 & = & -a\omega G^{a-1,n}_{1,0}+(n-a)\omega G^{a+1,n}_{1,0} -
\frac{U''(q)a}{m\omega} G^{a-1,n}_{1,0} +
\frac{U'''(q)a}{2(m\omega)^{3/2}}G^{0,2}_{0,0} G^{a-1,n-1}_{0,0} \nonumber\\
&& \: - \frac{U'''(q)a}{2(m\omega)^{3/2}} \left(G^{a-1,n+1}_{0,0} -
\frac{(a-1)(a-2)}{12} G^{a-3,n-3}_{0,0}\right)\,.
\label{EOM}
\end{eqnarray}
The solutions at $O(\hbar^0)$ are given by \eqref{00solution} for even $a$ and
$n$, and $G^{a,n}_{0,0}=0$ for odd $a$ or $n$.  We can use this result in
\eqref{EOM} to obtain solutions for the $G^{a,n}_{1,0}$.  For even $a$ or
$n$ in \eqref{EOM}, all the $G_{0,0}$ terms vanish, and the equation is
identical to \eqref{18}, giving the same solution for the moments involved:
\begin{eqnarray}
G^{a,n}_{1,0} & = & 0 \ \text{for odd} \ a \label{odda} \\ && \:
\nonumber \\
G^{a,n}_{1,0} & = & C_n''\frac{(n-a)!a!}{((n-a)/2)!(a/2)!}
\left(1+\frac{U''(q)}{m\omega^2}\right)^{(2a-n)/4} \ \text{for even} \ a \ \text{and} \ n
\label{10eveneven}
\end{eqnarray}
As before at first adiabatic order, we implement the ground-state condition by
requiring $G_{0,0}^{a,n}+\sqrt{\hbar} G_{1,0}^{a,n}$ to agree with the known
harmonic values when $U(q)=0$; thus, $C_n+\sqrt{\hbar}C_n''=2^{-n}$ ($C_n''$
is not dimensionless). Again keeping $G_{0,0}^{a,n}$ unchanged compared to
(\ref{00solution}), we have $C_n''=0$ and therefore $G_{1,0}^{a,n}=0$ for even
$a$ and $n$.

For odd $a$ and $n$ in \eqref{EOM}, the $G_{0,0}$ terms do not vanish.
Substituting \eqref{00solution} into \eqref{EOM} and simplifying the resulting
expression, we have
\begin{eqnarray}
0 & = &
(n-a)G^{a+1,n}_{1,0}-a\left(1+\frac{U''(q)}{m\omega^2}\right)G^{a-1,n}_{1,0}
\nonumber\\ && \: +
\frac{U'''(q)a}{m^{3/2}\omega^{5/2}}\frac{4a-3n-1}{12\pi}\Gamma
\left({\frac{a}{2}}\right)\Gamma \left({\frac{n-a+1}{2}}\right)
\left(1+\frac{U''(q)}{m\omega^2}\right)^{(2a-n-3)/4}  \,.
\label{EOMsimplified}
\end{eqnarray}
For the case $a=n$, this gives
\begin{equation}
G^{n-1,n}_{1,0}=\frac{U'''(q)a}{m^{3/2}\omega^{5/2}}\frac{n-1}{12\pi}\Gamma
\left({\frac{n}{2}}\right)\Gamma
\left({\frac{1}{2}}\right)\left(1+\frac{U''(q)}{m\omega^2}\right)^{(n-7)/4}  \,.
\label{n=a}
\end{equation}
We can plug this solution into the $a=n-2$ equation to solve for
$G^{n-3,n}_{1,0}$, and so on.  In general,
\begin{eqnarray}
G^{a-1,n}_{1,0} & = &
\frac{n-a}{a}\left(1+\frac{U''(q)}{m\omega^2}\right)^{-1}G^{a+1,n}_{1,0}
\nonumber\\ && \: +
\frac{U'''(q)}{m^{3/2}\omega^{5/2}}\frac{4a-3n-1}{12\pi}\Gamma
\left({\frac{a}{2}}\right)\Gamma\left({\frac{n-a+1}{2}}\right)
\left(1+\frac{U''(q)}{m\omega^2}\right)^{(2a-n-7)/4}  \,.
\end{eqnarray}

From \eqref{n=a}, we see that the two terms have the same power in
$1+U''(q)/m\omega^2$, which allows us to write the solution as
\begin{equation}
G^{a,n}_{1,0}=D_{a,n}\frac{U'''(q)}{m^{3/2}\omega^{5/2}}
\left(1+\frac{U''(q)}{m\omega^2}\right)^{(2a-n-5)/4}
\ \text{for even} \ a \ \text{and odd} \ n,
\label{10evenodd}
\end{equation}
where
\begin{eqnarray}
 D_{a,n}&=& \frac{(-1)^b\Gamma \left( \frac{n}{2}\right)}{12\pi
   (1-\frac{n}{2})_b}\Big((n-1)b!\sqrt{\pi}+(n-8b-1)
 \Gamma\left(b+\frac{1}{2}\right) \nonumber\\
&&- \sum_{c=0}^{b-2}(-1)^c (n-8(b-c-1)-1)
\Gamma\left(b-c-\frac{1}{2}\right)(-b)_{c+1} \Big)
\end{eqnarray}
if $n\geq5$ and $b\geq2$, and
\begin{eqnarray}
D_{a,n} = \left\{
\begin{array}{rl}
\frac{n-1}{12\pi}\Gamma\left(\frac{n}{2}\right)\Gamma\left(\frac{1}{2}\right)
\ \ \ \ \ \ \ \text{if } n\geq3, \ b=0 & \\
\frac{3n-11}{12\pi(n-2)}\Gamma\left(\frac{n}{2}\right)\Gamma\left(\frac{1}{2}\right)
\ \ \ \ \ \ \ \text{if } n\geq3, \ b=1 & \\
\end{array} \right.
\label{D_an}
\end{eqnarray}
is a dimensionless prefactor that depends on $a$ and $n$.  In the above
expression, $b=(n-a-1)/2$ and $(x)_n=x(x+1)...(x+n-1)$ is the Pochhammer
symbol.
Comparing to \eqref{00solution}, we see that for odd $n$ and even $a$,
\begin{equation}
G^{a,n}_{1,0} \propto
\left(1+\frac{U''(q)}{m\omega^2}\right)^{-1}G^{a,n+1}_{0,0}\,.
\end{equation}
The additional dimensionful factor of $U'''/m^{3/2}\omega^{5/3}$ in
(\ref{10evenodd}) provides the correct dimension of $\hbar^{-1/2}$.

\subsubsection{First Adiabatic Order}

At first order in the adiabatic approximation and at $O(\sqrt{\hbar})$, the
equation of motion is
\begin{eqnarray}
\dot{G}^{a,n}_{1,0} & = & (n-a)\omega G^{a+1,n}_{1,1}-a\omega
\left(1+\frac{U''(q)}{m\omega^2}\right) G^{a-1,n}_{1,1} +
\frac{U'''(q)a}{2(m\omega)^{3/2}}\bigg(G^{0,2}_{0,0} G^{a-1,n-1}_{0,1}
\nonumber\\ && +   G^{0,2}_{0,1} G^{a-1,n-1}_{0,0} - G^{a-1,n+1}_{0,1} +
\frac{(a-1)(a-2)}{12} G^{a-3,n-3}_{0,1}\bigg)\,.
\label{EOM2}
\end{eqnarray}
If $a$ is odd, we see from \eqref{odda} that the left-hand side is zero.  If
$n$ is even, we see from \eqref{oddn} that the last four terms on the
right-hand side vanish.  Let us consider the simplest case first: odd $a$ and
even $n$.  In this case the equation is identical to \eqref{18}, giving the
same solution (\ref{00solutionC}) for the moments involved, with a new (and
dimensionfull) $C_n$. Recalling that the
$O(\hbar^{1/2}\lambda^0)$ and $O(\hbar^0\lambda)$ solutions \eqref{10eveneven}
and \eqref{01eveneven} were also the same, we can write
\[
G^{a,n}_{1,1}  \propto  G^{a,n}_{0,1}\propto G^{a,n}_{1,0}\propto
G^{a,n}_{0,0}
\]
Now requiring $G_{0,0}^{a,n}+G_{0,1}^{a,n}+\sqrt{\hbar}
(G_{1,0}^{a,n}+G_{1,1}^{a,n})$ to agree with the known ground-state moments in
the harmonic limit and keeping $G_{0,0}^{a,n}$ as in (\ref{00solution}), we
have
\begin{equation} \label{11eveneven}
G^{a,n}_{1,1} = G^{a,n}_{0,1}= G^{a,n}_{1,0}=0 \quad\mbox{for even }a\mbox{
  and }n\,.
\end{equation}

In these equalities we have made use of another relation.  It turns
out that the simplification of the equation of motion observed here is
by itself not sufficient to guarantee the same solution, because the
equation does not fully determine all the moments.  As in the
zeroth-order approximation in both $\sqrt{\hbar}$ and $\lambda$, a
constraint (\ref{constraint}) is needed from the next adiabatic order
in order to determine $G^{0,n}$ \cite{Karpacz}. (More specifically,
the constraint shows that $C_n$ does not depend on $q$.)  Here, the
second-order adiabatic equation of motion at $O(\sqrt{\hbar})$ is
\begin{eqnarray}
&& \dot{G}^{a,n}_{1,1} = (n-a)\omega G^{a+1,n}_{1,2}-a\omega
\left(1+\frac{U''(q)}{m\omega^2}\right) G^{a-1,n}_{1,2} +
\frac{U'''(q)a}{2(m\omega)^{3/2}}\bigg(G^{0,2}_{0,0} G^{a-1,n-1}_{0,2}
\nonumber\\ && + G^{0,2}_{0,1} G^{a-1,n-1}_{0,1} + G^{0,2}_{0,2}
G^{a-1,n-1}_{0,0} - G^{a-1,n+1}_{0,2} + \frac{(a-1)(a-2)}{12}
G^{a-3,n-3}_{0,2}\bigg)
\label{12EOM}
\end{eqnarray}
But $n$ is still even, so the last five terms vanish, again due to
\eqref{oddn}.  The right-hand side once again is the same for all cases
considered for now, so the same condition on the left-hand side follows as in
\cite{Karpacz}, and \eqref{11eveneven} is indeed correct.  We note, however,
that not all these equalities between the moments will be valid at second
order in $\lambda$ because $\dot{G}^{a,n}_{1,1}$, appearing at the left-hand
side of the equation of motion \eqref{12EOM}, will no longer be zero for odd
$a$ and even $n$.

Now let us consider the case where $a$ and $n$ in \eqref{EOM2} are
both even.  The extra terms on the right-hand side vanish, but the
left-hand side is given by the time derivative of $G_{1,0}^{a,n}$.
The resulting equation, which describes moments with odd $a$ and even
$n$, is identical to \eqref{01EOM}, the corresponding equation at
$O(\hbar^0)$ --- except that the coefficient $2^{-n}$ is the $C_n$
belonging to $G_{1,0}^{a,n}$, with $G_{1,0}^{a,n}=0$ for even $a$ and
$n$.  In this case the equation of motion fully determines all the
moments; no constraint from the next order references to harmonic
states are needed.  Consequently, the solution is
\begin{equation}
G^{a,n}_{1,1}=0 \ \ \text{for odd} \ a \ \text{and even} \ n.
\label{11oddeven}
\end{equation}

For odd $n$, the solutions are not the same as the $O(\hbar^0)$ solutions
\eqref{oddn}, which are zero.  In the case where $a$ and $n$ are both odd in
\eqref{EOM2}, the extra terms remain, but the left-hand side is zero. If we
compare \eqref{EOM2} to \eqref{EOM} and recall \eqref{11eveneven}, we see that
the additional terms are identical, except that the $G^{0,2}G^{a-1,n-1}$ term
in \eqref{EOM} effectively occurs twice in \eqref{EOM2}.  This has the effect
of replacing the factor of $4a-3n-1$ in \eqref{EOMsimplified} with $4a-3n+2$,
so the solution is the same up to a change in the prefactor in
\eqref{10evenodd}.

The final case, where $a$ is even and $n$ is odd in \eqref{EOM2} is more
difficult. The left-hand side is nonzero and given by the time derivative of
\eqref{10evenodd}, and the extra terms on the right-hand side involve the
expression given in \eqref{01oddeven}.  Here we will not attempt to find a
general solution.

\subsection{Adiabatic Approximation for $n=2$-moments at
  $O(\hbar)$}

At second order in $\sqrt{\hbar}$, the equations again get more
complicated. We will restrict ourselves here to deriving a relation
for solutions of second-order moments ($n=2$) as needed for the
leading correction in the equation of motion (\ref{pdot}). Our
considerations in this section only illustrate the procedure but do
not provide complete solutions.

The $n=2$-moments, which are the same at $O(\hbar^0)$ and $O(\hbar^{1/2})$,
are different at $O(\hbar)$.  The $O(\hbar)$ equation of motion for the
moments is given by \cite{EffAc}
\begin{eqnarray}
\dot { G}^{a,n}_2&=&-a\omega G_2^{a-1,n}
+(n-a)\omega G_2^{a+1,n}  -\frac{U''(q)a}{m\omega}G_2^{a-1,n}
\nonumber\\
&&+\frac{U'''(q)a}{2(m\omega)^{3/2}}\left( G_1^{0,2}
  G_0^{a-1,n-1}+ G_0^{0,2}
G_1^{a-1,n-1}\right)\nonumber\\
&&-\frac{U'''(q)a}{2(m\omega)^{3/2}}
\left(G_1^{a-1,n+1}-\frac{(a-1)(a-2)}{12}
G_1^{a-3,n-3}\right)
\nonumber\\
&&+\frac{U''''(q)a}{3!(m\omega)^2}
G_0^{0,3} G_0^{a-1,n-1}-\frac{U''''(q)a}{6(m\omega)^2}\left(
G_0^{a-1,n+2}-\frac{(a-1)(a-2)}{4}G_0^{a-3,n-2}\right)\,.
\end{eqnarray}
At zeroth order in the adiabatic approximation, the left-hand side is zero.
For $a=0$, we find that $G^{1,n}_{2,0}=0$, as at $O(\hbar^{1/2})$.  In
particular, $G^{1,2}_{2,0}=0$.  The $a=2$-equation gives no new information,
but confirms that $G^{1,3}_{1,0}=0$, as we found in Sec.~\ref{sec:10}.  The
$a=1$-equation is
\begin{equation}
0=\omega G^{2,2}_{2,0}-\omega \left(1+\frac{U''(q)}{m\omega^2}\right)
G^{0,2}_{2,0}-\frac{U'''(q)}{2(m\omega)^{3/2}}
G^{0,3}_{1,0}-\frac{U''''(q)}{6(m\omega)^2}G^{0,4}_{0,0} \,.
\end{equation}
The coefficients $G^{0,3}_{1,0}$ and $G^{0,4}_{0,0}$ are given by equations
\eqref{10evenodd} and \eqref{00solution}, respectively, so we can express
$G^{2,2}_{2,0}$ in terms of $G^{0,2}_{2,0}$:
\begin{equation}
G^{2,2}_{2,0}=\left(1-\frac{U''(q)}{m\omega^2}\right)G^{0,2}_{2,0}-
\frac{(U'''(q))^2}{24m^3\omega^5}
\left(1-\frac{U''(q)}{m\omega^2}\right)^{-2}+ \frac{U''''(q)}{8m^2\omega^3}
\left(1-\frac{U''(q)}{m\omega^2}\right)^{-1} \,.
\end{equation}
In order to find $G^{0,2}_{2,0}$, we would have to go to first order in $\lambda$.

\section{Equation of motion for the oscillator up to $O({\hbar}^{3/2})$ and the fourth adiabatic order}
\label{s:EOM}

It is evident from (\ref{G0}) compared to the previous equations that higher
time derivatives appear at higher orders of the adiabatic expansion. This
pattern continues at higher adiabatic orders, even going beyond the second
derivative order of the classical equations. Genuine higher-derivative
equations are then obtained when solutions for moments are inserted in
(\ref{Gdot}). Not surprisingly, it becomes more and more complicated to find
explicit solutions to higher orders, valid for generic $a$ and $n$. Still,
individual moments, such as $G^{0,2}$ as needed for the first corrections in
(\ref{pdot}), can be computed more easily because specific numbers take the
place of coefficients such as $A_n'$ and $B_n'$ subject to complicated
recurrence relations.

\subsection{Higher-derivative equation of motion}

From the preceding section it is clear how such calculations are
organized, and it suffices here to quote the results needed for the
leading corrections in the equations of motion of the anharmonic
oscillator, given as before by
\begin{eqnarray}
\nonumber  \dot{q}&=& m^{-1}p\label{eomqp}\\
\dot{p}&=&-m\omega^2q -U'(q)-\sum_{n=2}^{\infty}
\frac{1}{n!}\left(\frac{1}{m\omega
\hbar}\right)^{n/2}U^{(n+1)}(q)G^{0,n}\,.
\end{eqnarray}
Taking the time derivative of the $\dot{q}$ equation, we may write, correct up
to $O({\hbar}^{3/2})$ in quantum corrections,
\begin{eqnarray}
\ddot{q}=-\omega^2q -U'(q)/m
-\frac{\hbar}{2m^2\omega}U'''(q)\left(\sum_{\lambda}G^{0,2}_{0,\lambda}+
\sqrt{\hbar}\sum_{\lambda}G^{0,2}_{1,\lambda}\right) \,,
\label{EOMq}
\end{eqnarray}
showing which moments and orders we need.  (Moments of orders higher
than $n=2$ would be required at the next order, $O(\hbar^2)$.)  Here
we have already used the fact that $G^{0,3}_{0,i}=0$ for any value of
$i$, according to \eqref{oddn}.

In order to evaluate \eqref{eomqp} completely, we need to compute $G^{0,2}$ to
orders $O(\hbar^0)$ and $O(\hbar^{1/2})$ in the semiclassical expansion. The
previous section contains results up to the second adiabatic order, at which
the right-hand side of (\ref{EOMq}) would be of second order in time
derivatives. In particular, we have
\begin{equation}
 G^{0,2}_{0,0}=
\frac{1}{2}\left(1+\frac{U''(q)}{m\omega^2}\right)^{-1/2}
\end{equation}
from (\ref{00solution}), while
\begin{equation}
 G^{0,2}_{1,1}=G^{0,2}_{1,0}=G^{0,2}_{0,1}=0\,.
\end{equation}
At second adiabatic order, we have
\begin{equation}
G^{0,2}_{0,2}=\frac{U'''(q)\ddot{q}+U''''(q)\dot{q}^2}{16m\omega^4}
\left(1+\frac{U''(q)}{m\omega^2}\right)^{-5/2}
-\frac{5(U'''(q)\dot{q})^2}{64m^2\omega^6}
\left(1+\frac{U''(q)}{m\omega^2}\right)^{-7/2}
\end{equation}
while
\begin{equation}
 G_{1,2}^{0,2}=0\,.
\end{equation}

Higher adiabatic orders cannot be obtained from the previous general
formulas for $G^{a,n}$, but the relevant contributions to the specific
moment $G^{0.2}$ can be computed with the previous methods. Up to
fourth adiabatic order, we have $G_{0,3}^{0,2}=0$ (which holds for all
moments of even $a$ and $n$) by using (\ref{G0102}) at higher
adiabatic order and observing that $G_{0,2}^{a,n}=0$ for odd $a$ and
even $n$. Moreover, $G_{1,3}^{0,2}=0$ and $G_{1,4}^{0,2}=0$
vanish. The final moment we need, $G_{0,4}^{0,2}$ is more difficult to
derive, and we just sketch the procedure we followed. The fourth-order
equations can be manipulated to show a relation of the form
\begin{equation} \label{Deriv1}
G_{0,4}^{2,2}= X G_{0,4}^{0,2}+\frac{\Theta}{w}
\end{equation}
where $X=1+U''(q)/m\omega^2$ and
\begin{eqnarray}
\Theta &=&X^{-5/2}\left(\frac{1}{32m\omega^5}\right)\left[U'''(q)\ddddot{q}+4U''''(q)
\dddot{q}\dot{q}+3U''''(q)\ddot{q}^2+6U'''''(q)\dot{q}^2\ddot{q}+U''''''(q)
\dot{q}^4\right] \nonumber\\
&&-X^{-7/2}\left\{\left(\frac{5}{32m^2\omega^7}\right)\left[U''''(q)\dot{q}^2+
U'''(q)\ddot{q}\right]^2\right.\nonumber\\
&& \qquad \left.+\left(\frac{15}{64m^2\omega^7}\right)\left[U'''(q)\dot{q}\right]
\left[U'''(q)\dddot{q}+U'''''(q)\dot{q}^3+3U''''(q)\dot{q}\ddot{q}\right]\right\}\nonumber\\
&&+X^{-9/2}\left(\frac{245}{256m^3\omega^9}\right)\left[U'''(q)\dot{q}\right]^2
\left[U'''(q)\ddot{q}+U''''(q)\dot{q}^2\right] \nonumber\\
&&-X^{11/2}\left(\frac{315}{512m^4\omega^11}\right)\left[U'''(q)\dot{q}\right]^4
\end{eqnarray}
Then using the consistency equation (\ref{constraint}) from
\cite{EffAc}, we get
\begin{equation} \label{constraint4}
2X\dot{G}_{0,4}^{0,2}+\left(\frac{U'''(q)\dot{q}}{mw^2}\right)G_{0,4}^{0,2}+\frac{1}{\omega}
\dot\Theta=0 \,.
\end{equation}
Choosing $G_{0,4}^{0,2}=AX^{-7/2}+BX^{-9/2}+CX^{-11/2}+DX^{-13/2}$ as an
ansatz, we solve for these functions $A$, $B$, $C$ and $D$. However, we have
five equations (corresponding to the five different powers of $X$ in
(\ref{constraint4})) with four unknowns. This generates another consistency
equation which turns out to be satisfied by the coefficients in
\begin{eqnarray} \label{G4}
G^{0,2}_{0,4}&=&-\frac{U'''(q)\ddddot{q}+
4U''''(q)\dddot{q}\dot{q}+3U''''(q)\ddot{q}^2
+6U'''''(q)\dot{q}^2\ddot{q}+U''''''(q)\dot{q}^4}{64m\omega^6}
\left(1+\frac{U''(q)}{m\omega^2}\right)^{-7/2}\nonumber\\
&&+\left(\frac{21(U''''(q)\dot{q}^2+U'''(q)\ddot{q})^2}{256m^2\omega^8}\right.\nonumber\\
&&\qquad+\left.\frac{7(U'''(q)\dot{q})(U'''(q)\dddot{q}+3U''''\ddot{q}\dot{q}
+U'''''(q)\dot{q}^3)}{64m^2\omega^8}\right)
\left(1+\frac{U''(q)}{m\omega^2}\right)^{-9/2}\nonumber\\
&&-\frac{231(U'''(q)\dot{q})^2(U'''(q)\ddot{q}+U''''(q)\dot{q}^2)}{512m^3\omega^{10}}
\left(1+\frac{U''(q)}{m\omega^2}\right)^{-11/2}\nonumber\\
&&+\frac{1155(U'''(q)\dot{q})^4}{4096m^4\omega^{12}}
\left(1+\frac{U''(q)}{m\omega^2}\right)^{-13/2} \,.
\end{eqnarray}

We may now rewrite the equation of motion \eqref{eomqp} as:
\begin{eqnarray}
\ddot{q}&=&-\omega^2q-U'(q)/m\\
&&-\frac{\hbar}{2m^2\omega}  U'''(q)\left[f(q,\dot{q})+
f_1(q,\dot{q})\ddot{q}+f_2(q)\ddot{q}^2+f_3(q,\dot{q})\dddot{q}+
f_4(q)\ddddot{q}\right]\nonumber
\end{eqnarray}
where
\begin{eqnarray}
f(q,\dot{q})&=&\frac{1}{2}\left(1+\frac{U''(q)}{m\omega^2}\right)^{-1/2}+
\frac{U''''(q)\dot{q}^2}{16m\omega^4}\left(1+\frac{U''(q)}{m\omega^2}\right)^{-5/2}-
\frac{5(U'''(q))^2\dot{q}^2}{64m^2\omega^6}\left(1+\frac{U''(q)}{m\omega^2}\right)^{-7/2}\nonumber\\
&&-\frac{U''''''(q)\dot{q}^4}{64m\omega^6}\left(1+\frac{U''(q)}{m\omega^2}\right)^{-7/2}
+\frac{21(U''''(q))^2\dot{q}^4}{256m^2\omega^8}\left(1+\frac{U''(q)}{m\omega^2}\right)^{-9/2}\nonumber\\
&&+\frac{7U'''''(q)U'''(q)\dot{q}^4}{64m^2\omega^8}\left(1+\frac{U''(q)}{m\omega^2}\right)^{-9/2}
-\frac{231U''''(q)(U'''(q))^2\dot{q}^4}{512m^3\omega^{10}}\left(1+\frac{U''(q)}{m\omega^2}\right)^{-11/2}\nonumber\\
&&+\frac{1155(U'''(q))^4\dot{q}^4}{4096m^4\omega^{12}}\left(1+\frac{U''(q)}{m\omega^2}\right)^{-13/2}\,,
\end{eqnarray}
\begin{eqnarray}
f_1(q,\dot{q})&=&\frac{U'''(q)}{16m\omega^4}\left(1+\frac{U''(q)}{m\omega^2}\right)^{-5/2}
-\frac{3U'''''(q)\dot{q}^2}{32m\omega^6}\left(1+\frac{U''(q)}{m\omega^2}\right)^{-7/2}\nonumber\\
&&+\frac{63U''''(q)U'''(q)\dot{q}^2}{128m^2\omega^8}\left(1+\frac{U''(q)}{m\omega^2}\right)^{-9/2}
-\frac{231(U'''(q))^3\dot{q}^2}{512m^3\omega^{10}}\left(1+\frac{U''(q)}{m\omega^2}\right)^{-11/2}\,,
\end{eqnarray}
\begin{eqnarray}
f_2(q)&=&-\frac{3U''''(q)}{64m\omega^{6}}\left(1+\frac{U''(q)}{m\omega^2}\right)^{-7/2}
+\frac{21(U'''(q))^2}{256m^2\omega^{8}}\left(1+\frac{U''(q)}{m\omega^2}\right)^{-9/2}\,,\\
f_3(q,\dot{q})&=&-\frac{U''''(q)\dot{q}}{16m\omega^6}\left(1+\frac{U''(q)}{m\omega^2}\right)^{-7/2}
+\frac{7(U'''(q))^2\dot{q}}{64m^2\omega^8}\left(1+\frac{U''(q)}{m\omega^2}\right)^{-9/2}\,,\\
f_4(q)&=&-\frac{U'''(q)}{64m\omega^{6}}\left(1+\frac{U''(q)}{m\omega^2}\right)^{-7/2}
\end{eqnarray}
Once these coefficients are inserted in \eqref{eomqp} we have the
equation of motion for $q$ correct up to the fourth adiabatic order,
for quantum corrections up to $\hbar^{3/2}$. From (\ref{G4}), it is
now clear that higher time derivatives result, up to fourth order with
the present approximation. The equations for moments shown in this
section demonstrate that it is the adiabatic order, rather than the
semiclassical expansion or the order of moments, that determines the
order of derivatives. Although back-reaction of moments on expectation
values is responsible for higher time derivatives, the new degrees of
freedom that higher-derivative equations would imply if used at face
value, are not identical to the moments as true quantum degrees of
freedom.

\subsection{Uncertainty Relation and Zero-Point Energy}

Having derived solutions for second-order moments, we need to make sure that
they obey the uncertainty relation so that they can correspond to a
state. This requires us to compute not only the moment $G^{0,2}$, as used in
corrected equations of motion, but also $G^{1,2}$ and $G^{2,2}$. Since
expressions with these moments get more lengthy, we restrict the orders to
$(0,0)+(0,1)+(1,0)+(1,1)+(0,2)$. (Especially $G^{1,2}$ is more difficult to
obtain to higher orders.) We then have
\begin{eqnarray}
G^{0,2}&=&\frac{1}{2}\left(1+\frac{U''(q)}{m\omega^2}\right)^{-1/2}+
\frac{U'''(q)\ddot{q}+U''''(q)\dot{q}^2}{16m\omega^4}
\left(1+\frac{U''(q)}{m\omega^2}\right)^{-5/2}\nonumber\\
&-&\frac{5(U'''(q)\dot{q})^2}{64m^2\omega^6}
\left(1+\frac{U''(q)}{m\omega^2}\right)^{-7/2} \label{G02}\\
G^{2,2}&=&\frac{1}{2}\left(1+\frac{U''(q)}{m\omega^2}\right)^{1/2}-
\frac{U'''(q)\ddot{q}+U''''(q)\dot{q}^2}{16m\omega^4}
\left(1+\frac{U''(q)}{m\omega^2}\right)^{-3/2}\nonumber\\
&+&\frac{7(U'''(q)\dot{q})^2}{64m^2\omega^6}
\left(1+\frac{U''(q)}{m\omega^2}\right)^{-5/2}\label{G22}\\
G^{1,2}&=&-\frac{U'''(q)\dot{q}}{8m\omega^3}
\left(1+\frac{U''(q)}{m\omega^2}\right)^{-3/2}\,.
\end{eqnarray}

To check the uncertainty relation
$G^{0,2}G^{2,2}-(G^{1,2})^2\geq 1/4$, it is useful to
write $G^{0,2}=\frac{1}{2}X^{-1/2}+Y$ while
$G^{2,2}=\left(\frac{1}{2}X^{-1/2}-Y+\frac{(U'''(q)\dot{q})^2}
{32m^2\omega^6}X^{-7/2}\right)X$, where
$X=1+\frac{U''(q)}{m\omega^2}$ and
\[
Y=\frac{U'''(q)\ddot{q}+U''''(q)\dot{q}^2}{16m\omega^4}
\left(1+\frac{U''(q)}{m\omega^2}\right)^{-5/2}-
\frac{5(U'''(q)\dot{q})^2}{64m^2\omega^6}
\left(1+\frac{U''(q)}{m\omega^2}\right)^{-7/2} \,.
\]
With these definitions,  the required expression is of the form
\begin{eqnarray}
& &G^{0,2}G^{2,2}-(G^{1,2})^2\nonumber\\
&=&X\left[({\textstyle\frac{1}{4}}X^{-1}-Y^2)+({\textstyle\frac{1}{2}}X^{-1/2}+Y)
\frac{(U'''(q)\dot{q})^2}{32m^2\omega^6}X^{-7/2}-
\frac{(U'''(q)\dot{q})^2}{64m^2\omega^6}X^{-4}\right]\nonumber\\
&=&1/4-XY^2 +\frac{(U'''(q)\dot{q})^2}{32m^2\omega^6}X^{-5/2}Y\,. 
\label{CheckUncert}
\end{eqnarray}
In the limit $U(q)\rightarrow 0$, we have $X=1$ and $Y=0$, yielding a value of
$1/4$ for the above expression and exact saturation of the uncertainty
relation, in accordance with our assumption of the Gaussian ground state in
the harmonic limit. With anharmonicity, it is not obvious to see the sign of
$G^{0,2}G^{2,2}-(G^{1,2})^2-1/4$. It is, however, clear that we need $Y\geq
0$. The adiabaticity condition for the anharmonic ground state is therefore
not guaranteed to be valid automatically, but with our equations the
uncertainty relation can easily be monitored when numerical solutions are
analyzed.

Finally, having determined the moments $G^{0,2}$ as well as $G^{2,2}$ to some
orders, we can compute anharmonic corrections to the zero-point energy
$Z=\frac{1}{2}\hbar\omega (G^{0,2}+G^{2,2})=:\frac{1}{2}\hbar\omega Z'$.  With
(\ref{G02}) and (\ref{G22}) we have, now valid up to the order $(1,3)$,
\begin{eqnarray}
Z'&=&G^{0,2}+G^{2,2}\nonumber\\
&=&\frac{1}{2}\left(1+\frac{U''(q)}{m\omega^2}\right)^{-1/2}
+\frac{U'''(q)\ddot{q}+U''''(q)\dot{q}^2}{16m\omega^4}
\left(1+\frac{U''(q)}{m\omega^2}\right)^{-5/2}\nonumber\\
&&-\frac{5(U'''(q)\dot{q})^2}{64m^2\omega^6}
\left(1+\frac{U''(q)}{m\omega^2}\right)^{-7/2}\nonumber\\
&&+\frac{1}{2}\left(1+\frac{U''(q)}{m\omega^2}\right)^{1/2}
-
\frac{U'''(q)\ddot{q}+U''''(q)\dot{q}^2}{16m\omega^4}
\left(1+\frac{U''(q)}{m\omega^2}\right)^{-3/2}\nonumber\\
&&+\frac{7(U'''(q)\dot{q})^2}{64m^2\omega^6}
\left(1+\frac{U''(q)}{m\omega^2}\right)^{-5/2}\nonumber\\
&=&\frac{1}{2}X^{-1/2}(1+X)+Y(1-X)+\frac{(U'''(q)\dot{q})^2}
{32m^2\omega^6}X^{-5/2}\,.
\end{eqnarray}

\section{Conclusions}
\label{s:Conc}

We have shown explicitly how higher time derivatives arise in effective
equations of canonical quantum systems. Physically, quantum back-reaction by
moments of a state on expectation values implies correction terms, which in an
adiabatic approximation (combined with a semiclassical one) can be solved for
in terms of higher time derivatives. The quantum degrees of freedom are then
removed from direct view, analogous to integrating out variables in a path
integral. Although the quantum equations are local, given by differential
equations of finite order, the effective system in which infinitely many
degrees of freedom have been expressed by higher time derivatives becomes
non-local.

\subsection{General properties}

Our analysis has revealed several general properties of the expansions
considered.

First, some coefficients of moments vanish at all orders in one of the
expansions, irrespective of the harmonic state perturbed around. For instance,
as already mentioned, $G_{0,i}^{a,n}=0$ for odd $n$. This observation
simplifies some computations of moments relevant for corrected equations of
motion. The interesting moments that appear in quantum corrections, however,
satisfy equations that become progressively more involved as the orders
increase. The methods employed here can be extended to higher orders, but a
general solution seems out of reach.

Secondly, while coefficients of explicit solutions for moments are usually
complicated, the relationship between the adiabatic and derivative order is
evident from the general form of equations (\ref{Gdoti}). Going to higher
adiabatic orders requires the use of time derivatives of lower-order
coefficients, starting with zeroth adiabatic order in which coefficients
depend on $q$. In this way, higher and higher time derivatives enter solutions
as seen explicitly in the preceding section.

Thirdly, as in our explicit examples (\ref{00solutionC}),
(\ref{01oddeven}) and (\ref{32}), odd adiabatic orders mainly change
moments $G^{a,n}$ with odd $a$ (correlations), while even adiabatic
orders lead to corrections in moments with even $a$.

\subsection{Higher time derivatives in quantum cosmology}

Our results apply to all quantum systems, but are especially relevant for
quantum gravity and cosmology. In these settings, canonical techniques are
often crucial or at least applied widely, and it has remained unclear if and
how higher time derivatives should result. The present article shows this
unambiguously and provides a systematic procedure for their computation. As an
example for the importance of higher time derivatives, we may look at loop
quantum cosmology \cite{LivRev,Springer}. In this setting, one would
generically expect effective equations with higher time derivatives, as a
result of higher-curvature corrections. (We are dealing here with
quantizations of finite-dimensional systems, and must therefore leave
unaddressed the intriguing  possibility that strong quantum corrections may
lead to signature change from space-time to a 4-dimensional quantum version of
Euclidean space \cite{ScalarHol,Action,SigChange}. The question of how the
``evolution'' of states and quantum back-reaction can be formulated to compute
quantum effects in timeless Euclidean space requires further detailed
study. But quantum effective equations should still be non-local and require
higher derivatives, including spatial ones.)

However, higher derivative terms or the more general moment-dependent
corrections as in (\ref{pdot}) have not yet been computed in all
cases; adiabatic regimes may even be non-existing. (After all,
non-perturbative quantum gravity may lack a ground state with slowly
evolving moments.)  Instead, for a first impression of implications of
quantum effects one often ignored quantum back-reaction altogether,
considering only quantum-geometry corrections in homogeneous models
which are easier to implement by simple modifications of classical
equations. A prominent one is the so-called holonomy modification, by
which the Hubble term ${\cal H}^2=(\dot{a}/a)^2$ in the Friedmann
equation is replaced by a periodic function such as $\sin(\ell{\cal
H})^2/\ell^2$ with a length parameter $\ell$ (which could be the
Planck length); see e.g.\ \cite{EffHam,AmbigConstr}. This modification
is motivated by a property in the full theory of loop quantum gravity
\cite{Rov,ThomasRev}, according to which only holonomies but not the
gravitational connection can be represented as operators. A heuristic
interpretation often encountered, states that higher-order terms in an
expansion of $\sin(\ell{\cal H})^2/\ell^2= {\cal H}^2(1-\frac{1}{3}
\ell^2{\cal H}^2+O((\ell{\cal H})^4))$ are related to higher-curvature
terms. However, this interpretation overlooks the fact that
higher-curvature terms also provide higher time derivatives, which are
not included in most studies (and have not
yet been computed). But if the higher-derivative part of curvature
corrections is ignored, one cannot consider isolated higher powers of
spatial curvature components as a reliable expansion.  Generically,
there is no reason to assume that a term of ${\cal H}^2$ is more
important than, say, $\dot{{\cal H}}$, both of which contribute to the
space-time curvature scalar. Expansions become inconsistent when only
one type of terms is kept and, perhaps even more damningly, general
covariance is put in jeopardy.

With holonomy corrections, the whole series expansion of $\sin^2(\ell{\cal
  H})/\ell^2$ by $\ell{\cal H}$ is used, but not a single higher time
derivative. Such an approximation cannot be consistent unless one can show
that there are no higher time derivatives whatsoever. There is indeed a
harmonic system in loop quantum cosmology free of quantum back-reaction
\cite{BouncePert}, given by a spatially flat isotropic model with a free,
massless scalar. In this model, holonomy corrections correctly describe
quantum evolution. If one departs from this model just slightly, when matter
remains kinetic dominated and anisotropies and inhomogeneity are small,
quantum back-reaction does arise \cite{QuantumBounce,BounceSqueezed} but may
be assumed weak; holonomy corrections can still be reliable. In all other
cases, however, correct physical conclusions can be drawn only when all
quantum corrections have been computed. For instance, when there is a phase of
slow-roll inflation, the potential dominates and one must expect effective
equations based on the assumption of kinetic domination to break down. The
presence of higher-derivative terms means that holonomy corrections on their
own are not reliable, but it also suggests interesting relationships with
early-universe models based on higher-curvature or non-local derivative terms,
especially regarding the singularity issue
\cite{BounceDeriv,BounceDeriv2,BounceDerivStable}. There is still much work to
be done to apply the complete effective methods to quantum cosmology. At
the very least, the present article serves to clarify the role of
higher-derivative corrections.

\subsection{The Role of Adiabaticity}

The use of the adiabatic approximation as the key ingredient to arrive
at higher time derivatives implies that not all quantum regimes may be
amenable to a higher-derivative description. The validity of the
adiabatic approximation is an assumption, which can be tested
self-consistently but need not always be valid. We had to use
equations of different orders to determine all coefficients, and
depending on the given quantum dynamics, not all these equations may
be mutually consistent, for instance if anharmonic constructions are
attempted for fully squeezed, correlated harmonic coherent states,
contradicting the condition $G^{a,n}_{0,0}=0$ derived here for odd $a$
and even $n$. Moreover, the final solutions may be mathematically
consistent, but could violate the uncertainty relation. If they do,
there would be no state corresponding to the moment solutions. (See
\cite{BouncePot} for examples in quantum cosmology where these
problems occur.)  Solutions for moments describe a dynamical state,
and not all states may allow moments to evolve adiabatically. Here, we
have expanded around the ground state, which one can reasonably expect
to evolve slowly. In other, more excited states, the condition may not
be met.

If the adiabatic approximation does not apply, there are still
effective equations, and they can be expanded in a semiclassical
approximation. However, it is then no longer possible to solve for the
moments in terms of expectation values, and no higher-derivative
effective equations exist. One would rather work with a
higher-dimensional effective system, a dynamical system in which the
expectation values together with all moments relevant to a given order
in $\hbar$ are kept as in (\ref{pdot}) and (\ref{Gdot}). The
non-locality of the quantum system is then realized by the infinite
number of moments if all orders are included. To any finite order in
the moments, such a system can still be solved approximately, most
often by numerical means, and evaluated for physical information; see
e.g.\ \cite{HighDens,HigherMoments} for examples including a rather
large number of moments. Effective equations in terms of moments,
explicitly exhibiting the true quantum degrees of freedom, are
therefore more general than higher-derivative effective equations, but
higher-derivative equations, if they exist, can show some features
more directly.

\section*{Acknowledgements}

This work was supported in part by NSF grant PHY0748336.

\end{document}